\documentclass[aps,pra,reprint,showpacs,nofootinbib,superscriptaddress]{revtex4-1}
\usepackage{physics}
\usepackage{qcircuit}
\usepackage[super]{nth}
\usepackage{graphicx, comment}
\usepackage{amsfonts}
\usepackage{amsmath,amssymb}
\usepackage{bm}
\usepackage{color}
\usepackage{epstopdf}
\usepackage{epsfig}
\usepackage{subfigure}
\usepackage{verbatim}

\usepackage[colorlinks]{hyperref}
\AtBeginDocument{%
  \hypersetup{
    citecolor=blue,
    urlcolor=blue}}
\usepackage{lineno}
\usepackage[thicklines]{cancel}
\usepackage{color}
\usepackage{url}
\usepackage{dsfont}
\usepackage{float}
\graphicspath{ {./images/} } 
\usepackage{slashed}
\usepackage{nicefrac}

\begin{document}
\title{Quantum Computation of the Massive Thirring Model}

\date{\today}

\author{Chinmay Mishra}
\email{cmishra1@tennessee.edu}
\affiliation{Department of Physics and Astronomy,  The University of Tennessee, Knoxville, TN 37996-1200, USA}

\author{Shane Thompson}
\email{sthomp78@tennessee.edu}
\affiliation{Department of Physics and Astronomy,  The University of Tennessee, Knoxville, TN 37996-1200, USA}

\author{Raphael Pooser}
\email{pooserrc@ornl.gov}
\affiliation{Computational Sciences and Engineering Division, Oak Ridge National Laboratory,
  Oak Ridge, TN 37831, USA}
\affiliation{Department of Physics and Astronomy,  The University of Tennessee, Knoxville, TN 37996-1200, USA}
\author{George Siopsis}
\email{siopsis@tennessee.edu}
\affiliation{Department of Physics and Astronomy,  The University of Tennessee, Knoxville, TN 37996-1200, USA}

\begin{abstract}
    Relativistic fermionic field theories constitute the fundamental description of all observable matter.  The simplest of the models provide a useful, classically verifiable benchmark for noisy intermediate scale quantum computers.  We calculate the energy levels of the massive Thirring model -- a model of Dirac fermions with four-fermion interactions -- on a lattice in 1 + 1 space-time dimensions.  We employ a hybrid classical-quantum computation scheme to obtain the mass gap in this model for three spatial sites. With error mitigation the results are in good agreement with exact classical calculations. Our calculations extend to the vicinity of the massless limit where chiral symmetry emerges, however relative errors for quantum computations in this regime are significant. We compare our results with an analytical calculation using perturbation theory.
\end{abstract}

\maketitle

\section{Introduction}
The field of elementary particle physics represents a promising application area for quantum computers. While classical computers suffer from exponential growth of the Hilbert space for many-body systems, quantum computers have the potential to bypass this by exploiting superposition and entanglement \cite{Feynman1982}. Quantum field theory, consisting of infinite degrees of freedom, provides the framework for the Standard Model of particle physics. A standard way of calculating physical quantities in quantum field theory is by discretizing space-time and placing the system on a lattice. This enables a non-perturbative study of the Standard Model with large classical resource costs. For example, the machinery of lattice quantum chromodynamics (QCD), which is important for nuclear physics, employs the world's fastest supercomputers. Quantum computers promise to improve the speed of calculations in quantum field theories exponentially \cite{Jordan1130}.

Although work in this direction has been ongoing for more than a decade (see, for example, \cite{Byrnes2006}), the availability of noisy intermediate scale quantum (NISQ) computing hardware is relatively recent. This has enabled further work on quantum simulation of quantum field theories \cite{preskill_fermionic, martinez_real-time_2016, muschik_u1_2017, klco_quantum-classical_2018, Yeter-Aydeniz2018, Yeter-Aydeniz2019, Klco2019_digit,Klco2019,kokail_self-verifying_2019,2019arXiv191100003B, 2019arXiv190410440K}. Building off of techniques used for the quantum computation of $\phi^4$ scalar field theory \cite{Yeter-Aydeniz2019}, we performed a hybrid classical-quantum calculation of energy levels of the $1+1$ dimensional massive Thirring model \cite{Thirring1958}, which includes a four-particle interaction for fermions. We implemented the model simulation on IBM-Q's Boeblingen device. We provide multiple ans\"atze for calculating the ground and excited states, which range in width from two to six qubits. The complexity of the computation increases as we consider each ansatz in turn and as we move from the ground to excited state, making for a good QFT-based benchmark for near term devices.

Our discussion is organized as follows. In Section \ref{sec:2}, we introduce the massive Thirring model and place it on a spatial finite lattice. In Section \ref{sec:3}, we calculate energy levels analytically using perturbation theory, as well as by exact diagonalization. In Section \ref{algorithms}, we discuss multiple hybrid classical-quantum algorithms for computing the energy levels of the Hamiltonian, with the number of qubits employed varying from as few as two to as many as six. In Section \ref{sec:5}, we discuss error mitigation. In Section \ref{Results}, we present our results. In Section \ref{sec:7}, we offer concluding remarks. Appendix \ref{app:a} contains some details of our quantum computation.

\section{The Massive Thirring Model}
\label{sec:2}
    In this section, we describe the massive Thirring model and then proceed with the discretization of the same on a $1$-dimensional spatial lattice. We describe the changes to be incorporated as a result of the  discretization of the model. We end with a comment on an emerging chiral symmetry, which is a regime we later explore on quantum hardware.
\subsection{The Model}
The Thirring model is an exactly solvable model of interacting massless fermions in $1+1$ spacetime dimensions.  We consider the massive Thirring model. This model is also known to be equivalent to  sine-Gordon theory \cite{Coleman1975}. The Lagrangian for the model is 
\begin{equation}
\label{eq:model}
    \mathcal{L}
    = \bar{\psi} (i\gamma^\mu \partial_\mu - m_0) \psi + \frac{g^2}{4} \bar{\psi} \gamma^\mu \psi \bar{\psi} \gamma_\mu \psi~,
\end{equation}
where $\mu = 0,1$ and $\bar{\psi} = \psi^\dagger \gamma^0$. For the Dirac matrices, we use the representation
\begin{equation}
\gamma^0 = 
\left(
\begin{array}{cc}
0 & -i\\
i & 0
\end{array}
\right)\quad  \text{and} \quad \gamma^1= 
\left(
\begin{array}{cc}
0 & -i\\
-i & 0
\end{array}
\right)
\end{equation}
The system contains two parameters, $m_0$, which is the bare mass, and the dimensionless coupling constant $g^2$.

Notice that the conjugate momentum to the fermionic field is
\begin{equation}
    \pi = \frac{\partial\mathcal{L}}{\partial (\partial_0 \psi)} = i\psi^\dagger
\end{equation}
Therefore, the standard anti-commutation relations between a fermionic field and its conjugate momentum imply that the components of the fermionic field $\psi$ obey equal-time anti-commutation relations,
\begin{equation}\label{eq:3}
    [ \psi_\alpha^\dagger (x) , \psi_\beta (x') ]_+ = \delta_{\alpha\beta} \delta (x-x')
\end{equation}
with all other anti-commutators vanishing, where $\alpha,\beta = 1,2$ ($\psi$ consists of two components).

It is convenient to expand $\psi$ in terms of solutions of the Dirac equation $(i\gamma^\mu \partial_\mu - m_0 ) \psi = 0$ (which is the equation of motion of the non-interacting model) that form a complete set,
\begin{equation}\label{eq:fieldexpansion}
    \psi(x) = \int \frac{dk}{2\pi\sqrt{2\omega(k)}}\left[b(k) u(k) e^{ikx} + c^\dagger(k) v(k) e^{-ikx} \right]
\end{equation}
where $u(k) e^{ikx}$ ($v(k) e^{ikx}$) are plane-wave positive (negative) energy solutions of the Dirac equation. The operators $b^\dagger (k)$ ($c^\dagger (k)$) create a fermion (anti-fermion) of momentum $k$ and energy $\pm\omega(k)$, where $\omega(k) = \sqrt{k^2 + m_0^2}$. The anti-commutation relations \eqref{eq:3} for the fermion field $\psi$ imply the anti-commutation relations for the fermionic modes,
\begin{equation}\label{eq:6}
    [ b^\dagger (k) , b(k')]_+ = [c^\dagger (k) , c(k')]_+ = 2\pi\delta(k-k')
\end{equation}
with all other anti-commutators vanishing.

It is straightforward to obtain the Hamiltonian of the system,
\begin{equation}
    \mathcal{H} = -\bar{\psi} (i\gamma^1 \partial_1 - m_0) \psi - \frac{g^2}{4} \bar{\psi} \gamma^\mu \psi \bar{\psi} \gamma_\mu \psi
\end{equation}
We note that the massive Thirring model is equivalent to the Gross-Neveu model for a single fermionic species \cite{Gross1974}, which is known to be perturbatively renormalizable~\cite{Feldman1986}.

\subsection{Latticization of the Model}
To study the fermionic system on a digital computer, we must alter it in two ways: discretization of the spatial coordinate and introduction of the Wilson term in the Hamiltonian that vanishes in the continuum limit.

We  discretize space into $N$ sites, so that the spatial coordinate $x \rightarrow x_n \equiv na$, where $n \in \mathds{Z}_{N}$ is the site index and $a$ is the lattice spacing. Having a finite lattice spacing amounts to introducing an ultraviolet cutoff $\Lambda \sim \mathcal{O} ({1}/{a})$ in the system. The continuum limit is recovered by letting $a\rightarrow 0$ and $N\to\infty$. The other parameter in the system that has non-vanishing dimension (the bare mass $m_0$) can be replaced by the dimensionless combination $m_0a$. To simplify notation, we set the lattice spacing $a=1$, and denote the spatial sites by $x \in \mathds{Z}_N$. With this choice of units, the continuum limit is recovered as $N\to\infty$ and $m_0\rightarrow0$ (i.e., $m_0a \to 0$). 

The discretization implies a modification of the dependence of energy on momentum, which  would  otherwise be given by Einstein's dispersion relation. As is well-known, unlike in the case of scalar fields, this modification leads to the problem of fermion doubling. The latter can be resolved by incorporating the Wilson term into the Hamiltonian:
\begin{equation}\label{eq:8}
\mathcal{H}_\text{Wilson} = -\frac{\xi}{2}  \bar{\psi}_{x} {\nabla}^2\psi_x
\end{equation}
where the Wilson parameter $\xi$ is arbitrary ($0<\xi < 1$), and $\nabla$ denotes finite difference, $\nabla\psi_x = \psi_{x+1} - \psi_x$. The Wilson term vanishes in the continuum limit, as can be easily verified by restoring the lattice spacing $a$ and letting $a\to0$. 

Assuming periodic boundary conditions, the momentum $k$ is also discretized taking values on the dual lattice, $\frac{2\pi}{N} k$, where $k\in \mathds{Z}_N$ is the momentum quantum number.

The above modifications lead to the modified dispersion relation on the lattice,
\begin{equation} \label{eq:dispersion}
\omega_k = \sqrt{ \tilde{m}_k^2 +  \sin^2 \frac{2\pi k}{N}}~,~ \tilde{m}_k = m_0 + 2 \xi \sin^2 \frac{\pi k }{N}~.
\end{equation}
The expansion of the fermionic field \eqref{eq:fieldexpansion} is also modified to
\begin{equation}\label{eq:fieldexpansionl}
    \psi_x = \frac{1}{\sqrt{N}} \sum_{k=0}^{N-1} \frac{1}{\sqrt{2\omega_k}}  \left[b_k u_k e^{2\pi ikx/N}  + c_k^\dagger v_k e^{-2\pi ikx/N} \right]
\end{equation}
where
\begin{equation}
u_k = \left( \begin{array}{c}
\sqrt{\omega_k - \sin\frac{2\pi k}{N}}\\
i\sqrt{\omega_k + \sin\frac{2\pi k}{N}}
\end{array} \right)~,~
v_k = u_k^\ast
\end{equation}
The anti-commutation relations on the lattice read
\begin{equation}
    [b_k^\dagger , b_{k'} ]_+ = [c_k^\dagger , c_{k'} ]_+ = \delta_{kk'}
\end{equation}
where $b_k^\dagger$ ($c_k^\dagger$) creates a fermion (anti-fermion) with momentum quantum number $k$ (i.e., momentum $2\pi k/N$).

Expressing the Hamiltonian in terms of fermionic modes, we obtain
\begin{equation}\label{eq:13}
    H = H_0 + H_\text{int}
\end{equation}
where the non-interacting part is diagonalized,
\begin{equation}\label{eq:H0}
    H_0 = \sum_{k=0}^N \omega_k \left[ b_k^\dagger b_k + c_k^\dagger c_k \right]
\end{equation}
and the interacting part is
\begin{equation}\label{eq:Hint}
    H_\text{int} = \frac{g^2}{4} \sum_{x=0}^{N-1} \bar{\psi} \gamma^\mu \psi \bar{\psi} \gamma_\mu \psi
\end{equation}
and consists of 16 different terms if expressed in terms of fermionic modes. Each of these terms contains four modes which are mixtures of $b_k^\dagger, b_k , c_k^\dagger,$ and $c_k$.

An important physical quantity is the total fermion charge,
\begin{equation}
    Q_f = \sum_{k=0}^{N-1} \left( b_k^\dagger b_k - c_k^\dagger c_k \right)
\end{equation}
which counts the total number of fermions minus the number of antifermions in a state. It is a conserved charge, as is commutes with the Hamiltonian,
\begin{equation}
    [ H , Q_f ] = 0
\end{equation}
Therefore, the spectrum consists of states which are simultaneous eigenstates of the Hamiltonian and the total fermion charge. In the absence of interactions ($g=0$), the common eigenstates can be specified by the set of occupation numbers $\{ n_i\}, i=0,1,\dots, 2N-1$, where $n_i = 0,1$. We will order the occupation numbers so that an even (odd) site refers to a fermion (anti-fermion), i.e., $n_{2k} = 1$ ($n_{2k+1} =1$) means that the state contains a fermion (anti-fermion) of momentum quantum number $k$. In terms of modes,
\begin{equation}\label{eq:18}
    \ket{\{ n_i \} } = (b_0^\dagger)^{n_0} (c_0^\dagger)^{n_1} \cdots (b_0^\dagger)^{n_{2N-2}} (c_0^\dagger)^{n_{2N-1}} |\Omega\rangle
\end{equation}
where $\ket{\Omega}$ is the vacuum state annihilated by all modes,
\begin{equation}
    \{ b_k , c_k \} \ket{\Omega} = 0
\end{equation}
i.e, $\ket{\Omega} = \ket{00\cdots 0}$.

These states are eigenstates of both $H$ and $Q_f$, and have energy
\begin{equation}\label{eq:20}
    E_{\{ n_i \} }^{(0)} = \sum_{k=0}^{N-1} (n_{2k} + n_{2k+1} ) \omega_k
\end{equation}
and fermion quantum number
\begin{equation}
    Q_f = \sum_{k=0}^{N-1} (n_{2k} - n_{2k+1})
\end{equation}
For example, the vacuum state $\ket{\Omega}$ has vanishing energy and fermion quantum number. The state $\ket{100\ldots} = b_0^\dagger \ket{\Omega}$ has energy $\omega_0 = m_0$ and $Q_f = +1$, representing a single fermion at rest, whereas $\ket{010\ldots} = c_0^\dagger \ket{\Omega}$ also has energy $m_0$, but opposite $Q_f = -1$, representing a single anti-fermion at rest.

If the system is prepared in one of these eigenstates, and evolves under the Hamiltonian \eqref{eq:13} including interactions, transitions may occur between different states. The states $\ket{\{ n_i\} }$ are not energy eigenstates in the presence of interactions. However, the fermion quantum number is conserved during evolution which implies that fermions and anti-fermions are always created in pairs.

It should be noted that Thirring's original model is obtained by setting the bare mass parameter $m_0=0$. It exhibits an interesting feature, namely chiral symmetry. Indeed, it is easy to see that the Lagrangian in the continuum \eqref{eq:model} with $m_0=0$ is invariant under the chiral transformation $\psi \mapsto \gamma_5 \psi$, where $\gamma_5 = -\gamma^0 \gamma^1$. There are two obstructions to the study of this symmetry with the lattice Hamiltonian \eqref{eq:13}. One is the presence of zero modes if one sets $m_0=0$, because the frequency vanishes for $k=0$ ($\omega_0 =0$, if $m_0=0$ in eq.\ \eqref{eq:dispersion}). To remedy this, one can introduce an infrared cutoff by adding $1/N$ to the mass parameter \cite{Yeter-Aydeniz2018}. This cutoff vanishes in the continuum limit ($N\to\infty$). The other obstruction is due to addition of the Wilson term \eqref{eq:8}, which is forced upon us by the fermion doubling problem, but breaks chiral symmetry. The Wilson term vanishes in the continuum limit, but on a finite lattice one can only observe approximate chiral symmetry. Additionally, as we will show in subsequent sections, the quantum computation introduces errors which become more significant in the limit of chiral symmetry.

\section{Mass Gap}
\label{sec:3}

Next, we obtain the mass gap, which is a physical quantity, by calculating the energy levels of the Hamiltonian \eqref{eq:13} corresponding to the ground and first excited states, $E_0$ and $E_1$, respectively. The mass gap is given by
\begin{equation}\label{eq:22}
    m = E_1 - E_0
\end{equation}
We perform the calculation analytically using first-order perturbation theory, as well as by exact diagonalization of the Hamiltonian using the Jordan-Wigner transform. 
\subsection{Perturbative Analysis}

Using first-order perturbation theory, it is straightforward to obtain an analytic expression for the eigenstates of the Hamiltonian \eqref{eq:13} and the mass gap \eqref{eq:22}.

At zeroth order, the Hamiltonian is diagonal (eq.\ \eqref{eq:H0}). The ground state $|\Omega\rangle$ has zero energy ($E_0^{(0)} =0$), and the first excited state has zero momentum. Evidently, the latter has energy $E_1^{(0)} =\omega_0 = m_0$, and is doubly degenerate. The corresponding eigenstates are in the span of $b_0^\dagger |\Omega\rangle$ and $c_0^\dagger |\Omega\rangle$, which represent a fermion and an anti-fermion, respectively, at rest (momentum $k=0$). It follows that the mass gap at zeroth order is $m=m_0$ (bare mass). The mass gets renormalized due to quantum effects once interactions are turned on.

First-order corrections are governed by the interaction Hamiltonian \eqref{eq:Hint}, which consists of 16 different types of terms, as is evident from the mode expansion \eqref{eq:fieldexpansion} of the fermion field. Not all terms contribute at each energy level.

Only two of the 16 different terms in the interaction Hamiltonian contribute to the first-order correction to the ground-state energy level. We obtain
\begin{equation}
\delta E_0 = \bra{\Omega}H_\text{int}\ket{\Omega} \hspace{\fill}
= \frac{g^2}{2N} \left[ N^2 + \varepsilon_1^2 - \varepsilon_2^2  \right]
\end{equation}
where
\begin{equation}
    \varepsilon_1 = \sum_{k =0}^{N-1}  \frac{\tilde{m}_{k} }{\omega_k} \ , \ \ \varepsilon_2 = \sum_{k =0}^{N-1}  \frac{ \sin \frac{2\pi k}{N} }{\omega_k}
\end{equation}
The ground state, including first-order corrections, is
\begin{equation}
   \ket{\Omega} - \sum_{\{ n_i \} } \frac{\mathcal{T}_{\{ n_i \} , 0 } }{E_{\{ n_i \} }^{(0)} } \ket{{\{ n_i \} } } + \mathcal{O} (g^4)
\end{equation}
where we are summing over all combinations of occupation numbers that contain at least one non-vanishing number, and
\begin{equation}
    \mathcal{T}_{\{ n_i \} , 0 } = \bra{ \{ n_i \} } H_\text{int} \ket{\Omega}
\end{equation}
is the transition amplitude from the ground state of the free Hamiltonian $H_0$ to an excited state of energy $E_{\{ n_i \} }^{(0)}$ \eqref{eq:20}. Evidently, high-energy contibutions are suppressed. The most significant contributions are due to single fermion-antifermion pairs of opposite momenta. Denoting these states by $\ket{k,-k} = c_{N-k}^\dagger b_k^\dagger\ket{\Omega}$, we obtain the transition amplitudes

\begin{equation}
    \mathcal{T}_{\{ k,-k \} , 0} = \bra{k,-k} H_\text{int} \ket{\Omega}
    =\frac{g^2 \varepsilon_1}{2N} \frac{\sin{\frac{2\pi k}{N}}}{\omega_k}
\end{equation}
Only two of the 16 terms in the interaction Hamiltonian contribute to these amplitudes.

Since the $k=0$ transition amplitude vanishes, the largest contribution is due to $k=1, N-1$, each of energy $2\omega_1$ (creation of fermion--anti-fermion pair of opposite minimum momenta). Thus, the ground state is
\begin{equation}
    \ket{\Omega} - \frac{\mathcal{T}_{\{ 1,-1\} ,0}}{2\omega_1} \left(c_{N-1}^\dagger b_1^\dagger - c_1^\dagger b_{N-1}^\dagger \right) \ket{\Omega}+ \dots
\end{equation}
Working similarly for the first excited state, we note that only five of the 16 terms in the interaction Hamiltonian contribute. The correction to the energy level of the state $b_0^\dagger \ket{\Omega}$ (representing a single fermion at rest) is
\begin{equation}\label{eq:29}
\begin{aligned}
\delta E_1 &= \bra{\Omega} b_0 H_\text{int}b_0^\dagger \ket{\Omega}\\
&= \bra{\Omega} [b_0, H_\text{int} ] b_0^\dagger \ket{\Omega} + \delta E_0\\
&= -\frac{g^2}{2N}\varepsilon_1 + \delta E_0
\end{aligned}
\end{equation}
The correction to the energy level of the anti-fermion excitation is numerically the same (degenerate energy level).

The first excited states are found by reasoning as in the case of the ground state. For the first excited state with fermion quantum number $+1$, at first order we have three terms with fermion quantum number $+1$ each consisting of three particles of minimum total energy, $2\omega_1 + m_0$. We obtain

\begin{equation}
\begin{aligned}
    b_0^\dagger\ket{\Omega} &- \frac{\mathcal{T}_{\{ 1,-1\} ,1}}{2\omega_1 } \left( c_{N-1}^\dagger b_1^\dagger - c_1^\dagger b_{N-1}^\dagger\right) b_0^\dagger\ket{\Omega}\\
    &- \frac{\mathcal{T}'_{\{ 1,-1\} ,1}}{2\omega_1}  b_{N-1}^\dagger b_1^\dagger c_0^\dagger \ket{\Omega} + \dots
\end{aligned}
\end{equation}

where
\begin{align}
    \begin{split}
        \mathcal{T}_{\{ 1,-1\} ,1} &= \bra{\Omega} b_0 b_1 c_{N-1} H_\text{int} b_0^\dagger \ket{\Omega} \\ &= \frac{g^2}{2N}(\varepsilon_1-1)\frac{\sin\frac{2\pi}{N}}{\omega_1}
    \end{split}
\end{align}
is the transition amplitude for the creation of a fermion--anti-fermion pair of minimum non-vanishing momentum, and
\begin{equation}
    \mathcal{T}'_{\{ 1,-1\} ,1} = \bra{\Omega} c_0 b_1 b_{N-1} H_\text{int} b_0^\dagger \ket{\Omega} = \frac{g^2}{N}\frac{\sin\frac{2\pi}{N}}{\omega_1}
\end{equation}
is the transition amplitude to a state consisting of an anti-fermion at rest and a fermion pair with opposite minimum non-vanishing momenta. The three-particle system has the same fermion quantum number, $+1$, and total energy, $2\omega_1 + m_0$, as the other terms.

The corrections to the first-excited state of the anti-fermion excitation are calculated similarly.

Using Eq.\ \eqref{eq:29}, we obtain the first-order correction to the mass gap
\begin{equation}
    \delta m = \delta E_1 - \delta E_0 = -\frac{g^2}{2N}\varepsilon_1~.
    \label{pert_mass_gap}
\end{equation}
For a fixed $m_0$, as we increase the coupling constant $g^2$, the mass gap decreases, and eventually vanishes at a critical value $g_\text{crit}^2$. Notice that for large $m_0$, first-order perturbation theory yields
\begin{equation}\label{eq:34}
    g_\text{crit}^2 \approx 2m_0
\end{equation}
where we used $\varepsilon_1 \approx N$, for large $m_0$.

In the continuum limit ($N\to\infty$, $m_0,\xi\to 0$),
\begin{equation}
\begin{aligned}
    \delta m &= \frac{g^2}{2} \int_0^{2\pi} \frac{dk}{2\pi} \frac{m_0 + 2\xi \sin^2 \frac{k}{2}}{\sqrt{(m_0 + 2\xi\sin^2 \frac{k}{2})^2 + \sin^2 k}}\\
    &= \frac{g^2}{\pi} m_0 \log \frac{1}{m_0} + \dots
\end{aligned}
\end{equation}
showing that the fermion mass is renormalized to
\begin{equation}
    m = m_0+ \delta m = m_0 + \frac{g^2}{\pi} m_0 \log \frac{1}{m_0} + \dots
\end{equation}
%
Fig.\ \ref{exact_mass} shows a plot for the mass gap as a function of the coupling constant $g^2$ for $m_0 =10$ and $N=3$, along with exact (numerical) results from exact diagonalization of the Hamiltonian \eqref{eq:13}. Notice the striking agreement of perturbative and exact results (the small difference, occurring near the critical line where the mass gap vanishes, can be seen in Fig.\ \ref{exact_and_perturbative_m}) to be discussed next.

\subsection{Exact Diagonalization}
\label{diagonalization}

One can go beyond perturbation theory and obtain exact eigenvalues and corresponding eigenstates of the Hamiltonian \eqref{eq:13} including interactions by performing an exact diagonalization numerically. To this end, one needs to employ a representation of the fermionic modes $b_k$ and anti-fermionic modes $c_k$. We will use the Jordan-Wigner transform
\cite{Ortiz2001}, in which different modes are represented by Pauli matrices $X,Y,Z$. The Hilbert space is mapped onto a tensor product of $2N$ two-dimensional spaces, each spanned by the computational basis states, $\{ \ket{0}, \ket{1} \}$, which are eigenstates of $Z$ with corresponding eigenvalues $\pm 1$. We map $b_0 \mapsto X_0^-$, and
\begin{equation} \label{eq:JW}
    \begin{aligned}
    b_k &\mapsto \prod_{i=0}^{2k-1}(-Z_i)\ X_{2k}^- \ \ \ \ (k>0) \\
    c_k &\mapsto \prod_{i=0}^{2k}(-Z_i)\ X_{2k+1}^-
\end{aligned}
\end{equation}
where $X^\pm = X\pm iY$ and indices denote the position of the two-dimensional space in the tensor product on which the Pauli matrix acts.

The basis states \eqref{eq:18} are naturally mapped onto the computational basis states
\begin{equation}
    \ket{n_0\ n_1\ n_2\ \dots\ n_{2N-2}\ n_{2N-1}}
\end{equation}
where each entry can be 0 or 1, and $n_{2k}$ ($n_{2k+1}$) denotes the number of fermions (anti-fermions) of momentum quantum number $k$. For example, the ground state $\ket{\Omega}$ of the free Hamiltonian $H_0$ is mapped onto $\ket{000\dots0}$, and the first excited states $b_0^\dagger \ket{\Omega}$ (fermion excitation) and $c_0^\dagger \ket{\Omega}$ (anti-fermion excitation) are mapped onto $\ket{100\dots0}$ and $\ket{010\dots 0}$, respectively.

The resulting Hilbert space is $2^{2N}$-dimensional and the Hamiltonian is mapped onto a $2^{2N}\times 2^{2N}$ matrix. Diagonalizing the Hamiltonian for large $N$ is a daunting task. Moreover, expressing the Hamiltonian in terms of strings of Pauli matrices, which is needed for quantum computation, leads to a proliferation of terms. For example, even for $N=3$, we obtain 166 terms in the expression for the Hamiltonian once the interaction term is considered. Fortunately, most of the terms do not contribute at low energy levels leading to manageable expressions.

\begin{figure}[h]
    \centering
    \includegraphics[scale=0.45]{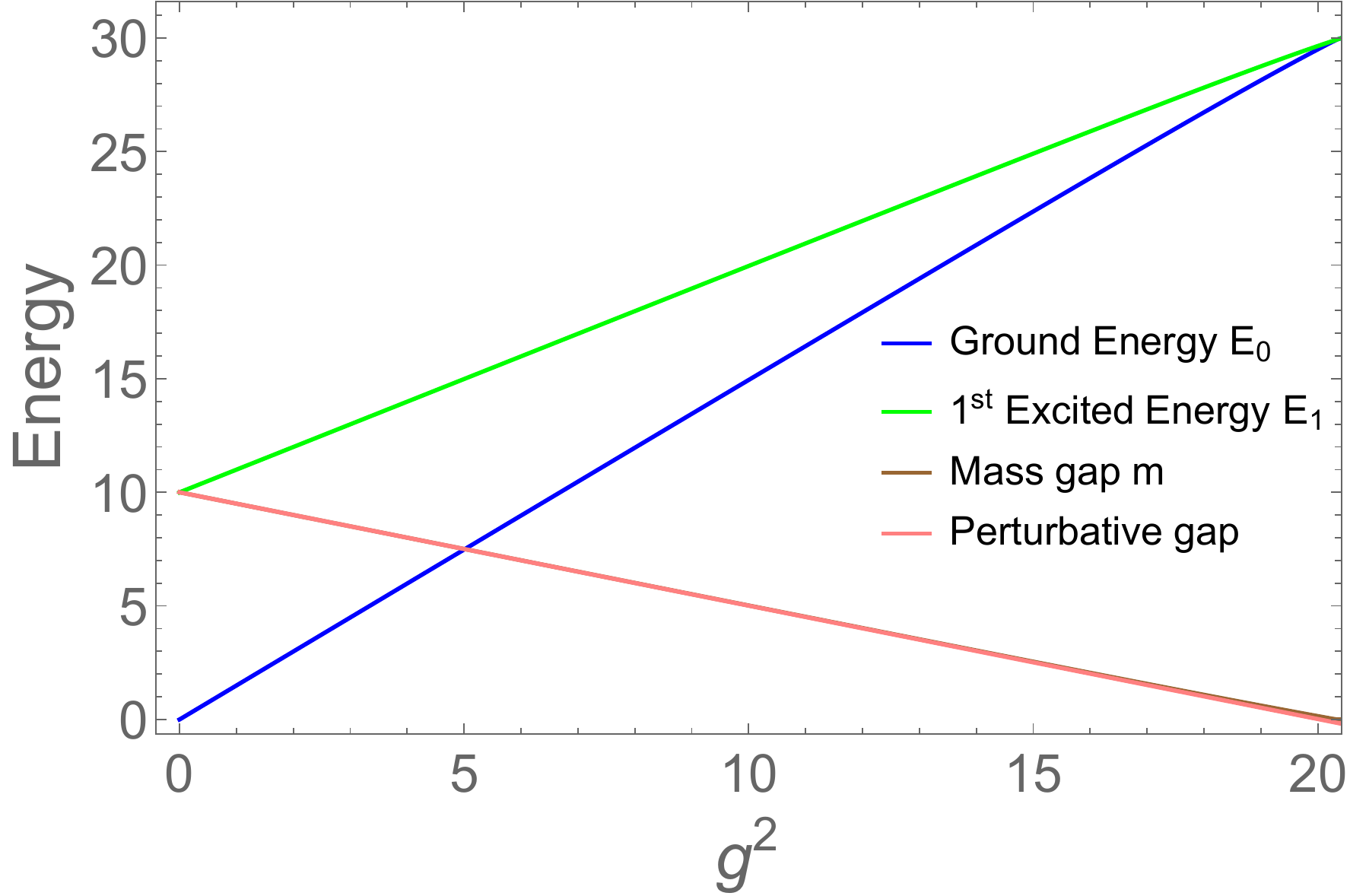}
    \caption{Ground-state and first-excited-state energies along with the mass gap (difference of these energies) as functions of interaction strength $g^2$, at $m_0=10$, $\xi =0.7 $, and $N=3$. The gap is computed from exact results as well as perturbation theory. The results are almost indistinguishable all the way to critical point ($m=0$).}
    \label{exact_mass}
\end{figure}

\begin{figure}[h]
    \centering
    \includegraphics[scale=0.45]{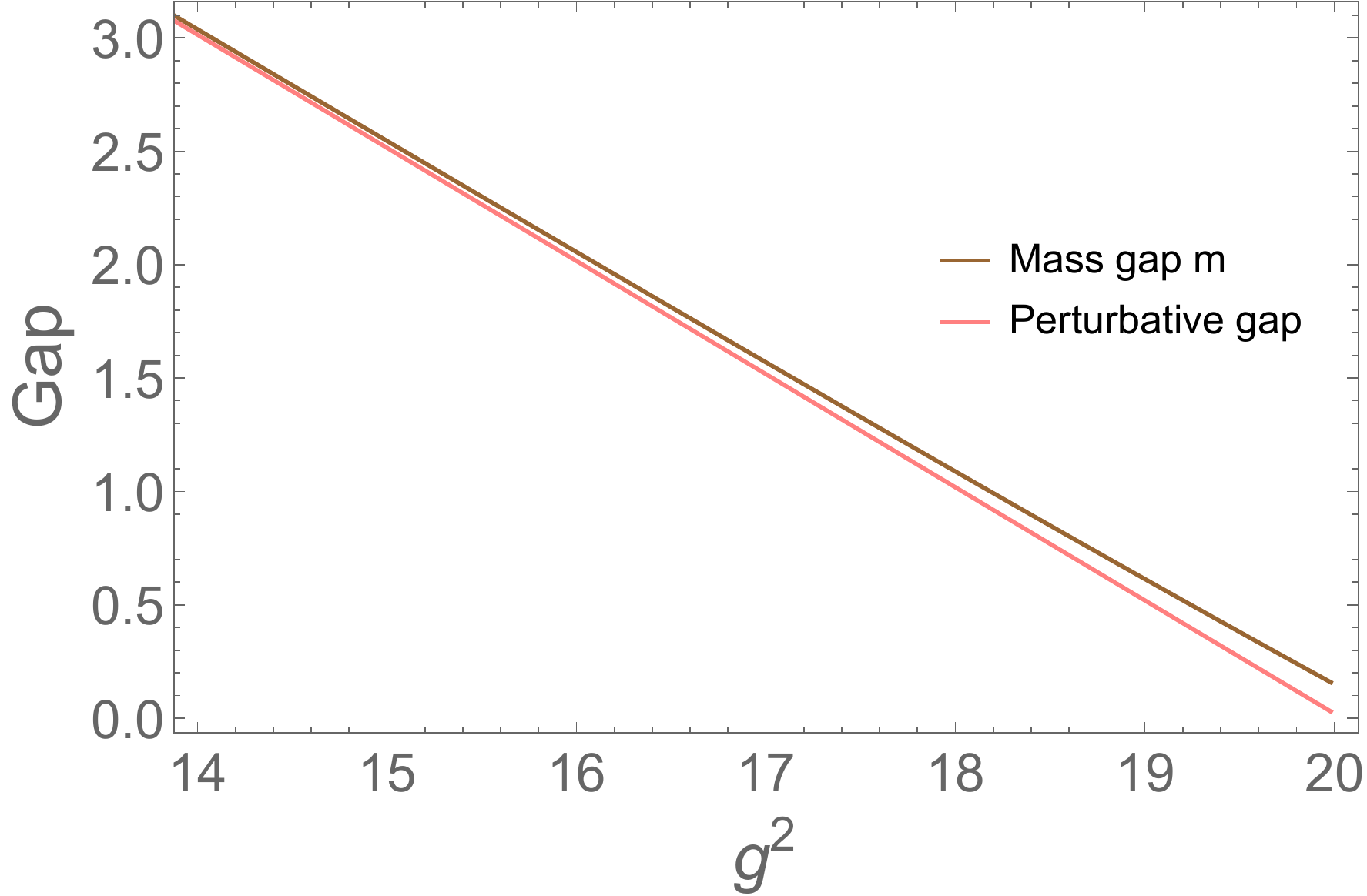}
    \caption{Mass gap $m$ vs interaction strength $g^2$ from Fig. \ref{exact_mass}, zoomed in to where perturbation theory diverges from exact results in the vicinity of the critical line where the mass gap vanishes.}
    \label{exact_and_perturbative_m}
\end{figure}

Results from exact diagonalization of the Hamiltonian are shown in Fig. \ref{exact_mass} for $N=3$ ($64\times 64$ matrix) and for the choice of parameters $m_0=10$, $\xi = 0.7$, and compared with analytic results from first-order perturbation theory \eqref{pert_mass_gap}. Analytic results for energy levels are in remarkable agreement with exact results owing to the fact that higher-order corrections in perturbation theory are suppressed for large bare mass parameter $m_0$. The small divergence of perturbative from exact results is shown in Fig. \ref{exact_and_perturbative_m} and occurs near the critical point.
\begin{figure}[h]
    \centering
    \subfigure{\includegraphics[width=0.4\textwidth]{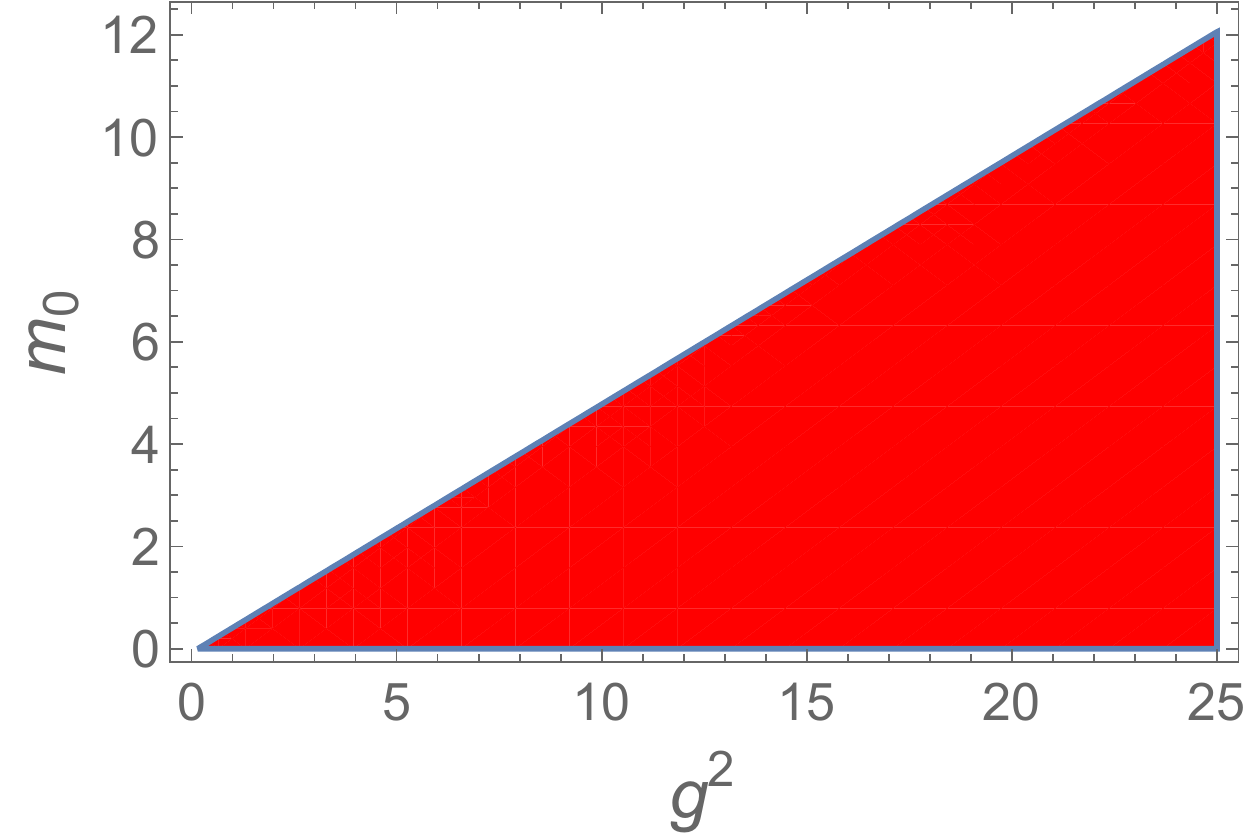}}
    \caption{The critical line (boundary of shaded region) where the mass gap vanishes in an $m_0$ vs.\ $g^2$ plot, at $\xi=0.7$ and $N=3$. For large $m_0$, the critical line agrees with the perturbative result \eqref{eq:34}.}
    \label{fig:3}
\end{figure}

Fig. \ref{fig:3} shows the critical line on which the mass gap vanishes. It agrees well with the asymptotic result \eqref{eq:34} from first-order perturbation theory. In the physically relevant region above the critical line, we expect good agreement between perturbative and exact results near the vertical axis (small coupling constant $g^2$) and also at large bare mass parameter $m_0$. This leaves only a small region where perturbation theory diverges from exact results. It turns out that, even in that region, the discrepancy is small. This makes perturbation theory a good guide towards quantum computation to which we turn our attention next.

\section{Quantum Computation} \label{algorithms}
In this section, we describe the  quantum computation of the ground and first-excited state energies using a hybrid classical-quantum algorithm based on the variational method. The mass gap follows from Eq.\ \eqref{eq:22}. To this end, one introduces a trial wavefunction $\ket{\psi_\text{trial}} = U\ket{00\dots0}$, where $U$ is a unitary that depends on certain variational parameters. One then estimates the energy by $\bra{\psi_{\text{trial}}} H \ket{\psi_\text{trial}} = \bra{0} U^\dagger H U\ket{0}$. For quantum computation, one expands the Hamiltonian $H = \sum_j H_j$, where each $H_j$ is unitary (a string of Pauli matrices), diagonalizes $H_j$ by writing $H_j = V_j^\dagger h_j V_j$, where $h_j$ is a string of Pauli $Z$ matrices, and forms quantum circuits by acting with $V_j U$ on $\ket{0} = \ket{00\dots0}$. This is the Unitary Coupled Cluster (UCC) method commonly used in quantum chemistry \cite{O'Malley2017}.

Here, we will introduce slightly more general Ans\"atze by writing the trial function as a superposition of trial functions that need two (or more) unitaries to specify them. This has the advantage of reducing the number of CNOT gates, which are significant sources of errors in NISQ quantum computers, at the expense of increasing the classical part of the computation. Another feature of this approach is that the size of the register required for a quantum computation will be reduced through such ans\"atze. Thus as we move away from such approaches in this section, we will see the computational complexity of the benchmark increase, along with errors from CNOT gates.

For quantum computations, we concentrate on the $N=3$ case which turns out to require, at most, four qubits for the ground state energy calculation and six for the first excited state energy. We write quantum registers as kets with all six qubits present and then present the simplified versions when less than six are required to run a circuit.
Results will be presented in Section \ref{Results}.

\subsection{Ground State} \label{GS_QC}

From perturbation theory, we know that the largest contributions to the ground state are from the states $\ket{000000}, \ket{0001100}, \ket{001001}$, all of which have vanishing total fermion quantum number. For $N=3$, these, together with $\ket{110000}$ whose contribution is expected to be suppressed (and confirmed by exact results), are the only states with vanishing fermion number and total momentum, so the ground state must be in their span. Due to charge conjugation symmetry (exchange of fermions and anti-fermions), the states $\ket{0001100}, \ket{001001}$ should contribute equally.

The above considerations inform the ansatz for the ground state
\begin{equation}\label{eq:39}
    \ket{\psi_\text{trial}}=a\ket{000000}+\frac{b}{\sqrt{2}} \left[ \ket{000110}+\ket{001001} \right]~,
\end{equation}
where $a,b\in\mathbb{R}$, and $a^2+b^2=1$. For $m_0 \gtrsim 1$, the ansatz \eqref{eq:39} produces a classical expectation value that differs from the true ground state energy by less than $0.001 \%$. In the quantum computation, we will work with two different equivalent expressions for the ground states following the above ansatz. The quantum circuits implementing these two expressions differ in the number of CNOT gates, allowing us to study the limitations of the quantum hardware after error mitigation. We will compute the first excited state similarly.

We note that the number of qubits required for the ground state computation is just four, since in all terms in \eqref{eq:39} the first two qubits are in the same state. The trial wavefunction can be implemented through the ansatz requiring a single CNOT gate,
\begin{equation} \label{gd_2term}
        \ket{\psi_\text{trial}}=\sin\theta\ U_1 \ket{000000} + \cos\theta\ U_2 \ket{000000}
        \end{equation}
        where
        \begin{equation}\label{eq:41} U_1 = e^{-i\frac{\phi}{2} X_3Y_4} \ , \ \ U_2 = X_2 X_5
\end{equation}
with variational parameters $\theta$ and $\phi$. 

$U_1$ and $U_2$ are implemented with the quantum circuits, 

\[\Qcircuit @C=1.75em @R=0.5em {
    \lstick{\ket{0}_3} & \qw & \targ & \qw  \\
    \lstick{\ket{0}_4} & \gate{R_Y (\phi)} & \ctrl{-1} & \qw  \\
}
\hspace{1.5cm}
\Qcircuit @C=0.5em @R=0.5em {
    \lstick{\ket{0}_2} & \qw & \gate{X} & \qw  \\
    \lstick{\ket{0}_5} & \qw & \gate{X} & \qw 
}\]
respectively, where $R_Y(\phi)$ implements a spinor a rotation around the $y$-axis of angle $\frac{\phi}{2}$. Note that these circuits are independent and implemented separately, and that each requires only two qubits for this particular ansatz.

For quantum computation, this is nontraditional in the sense that the two-term ansatz \eqref{gd_2term} is not given by a single unitary transformation acting on $\ket{0}$, as has been standard practice in quantum chemistry \cite{O'Malley2017}. Instead, the quantum computation  involves expectation values of the 
form
\begin{equation}
     \bra{0} U_m^\dagger H_j U_n \ket{0} \label{expectation_terms}
\end{equation}
where the $H_j$ are strings of Pauli matrices contributing to the Hamiltonian, and $m,n=1,\dots,t$, where t is the number of terms in the chosen ansatz ($t=2$ in our current case).

If $m=n$, we work in the standard way to diagonalize $H_j = V_j^\dagger h_j V_j$, where $h_j$ is diagonal in the computational basis (a string of Pauli $Z$ matrices). This is achieved by using 
\begin{equation}\label{eq:43}
    X = \mathbf{H} Z \mathbf{H} \ , \ \ Y = \mathbf{\mathcal{Y}}^\dagger Z \mathbf{\mathcal{Y}}
\end{equation}
where $\mathbf{H}$ is the Hadamard matrix and $\mathbf{\mathcal{Y}} = e^{i\frac{\pi}{4} X}$ \cite{Macridin2018}. Similar rotations can be used in the case $m\ne n$, except that in this case, we need to diagonalize the product $U_m^\dagger H_l U_n$, instead of just $H_l$, which may involve different rotations than those implemented by $\mathbf{H}$ and $\mathbf{\mathcal{Y}}$ (see appendix \ref{app:a} for details). Nevertheless, the diagonalization process is straightforward in the general case.

As mentioned above, the number of terms in the Hamiltonian is over 100, which requires unreasonably long access time given the queuing structure of today's quantum hardware. Fortunately, the number of terms can be reduced down to a handful in the case of the ground state by using standard simplifying techniques (see appendix \ref{app:a} for an outline).

The classical part of the algorithm can be significantly reduced by adopting an ansatz which consists of a single term, e.g., using 
\begin{equation}\label{eq:44}
        \ket{\psi_\text{trial}}={CX}_{25}X_2{CX}_{34}{CR_Y}_{23}(\phi)  {R_{Y}}_{2}(\theta)\ket{000000}
\end{equation}
with $\theta$ and $\phi$ as variational parameters,
and where ${CX}_{ij}$ is a CNOT gate with control qubit $i$ and target $j$, and ${CR_Y}_{ij}$ is a controlled rotation around the $y$-axis with control qubit $i$ and target $j$. As shown in the following circuit, the latter is implemented through a CNOT gate sandwiched between a rotation on the target by half the desired angle, and the inverse rotation \cite{nielsen_chuang_2010}. 
\[\Qcircuit @C=0.5em @R=0.5em {
    \lstick{\ket{0}_2} & \gate{R_Y(\theta)} & \ctrl{1} & \gate{X} & \qw & \ctrl{4} & \qw  \\
    \lstick{\ket{0}_3} & \gate{R_Y(\frac{\phi}{2})} & \targ & \gate{R_Y^\dagger(\frac{\phi}{2})} & \ctrl{1} & \qw &\qw  \\
    \lstick{\ket{0}_4} & \qw & \qw & \qw & \targ & \qw &\qw \\ \\
    \lstick{\ket{0}_5} & \qw & \qw & \qw & \qw & \targ &\qw
}\]
This simplifies the diagonalization process requiring only the standard diagonalization of the terms in the Hamiltonian, which can be achieved with $\mathbf{H}$, by writing $X = \mathbf{H} Z \mathbf{H}$. The price one pays for the simplification, in addition to the expansion of the register to four qubits, is the increase in the number of CNOT gates from one for the two-term ansatz \eqref{gd_2term} to three for the single-term ansatz \eqref{eq:44}, which increases two-qubit errors in NISQ quantum computers.

\subsection{First Excited State}

The first excited state corresponding to a fermion excitation receives contributions from computational basis states with fermion number $+1$. This guarantees that any ansatz will be orthogonal to the ground state. From perturbation theory, we are led to consider the ansatz
\begin{equation}\label{fetrial}
\begin{aligned}
    \ket{\psi_\text{trial}} &= a\ket{100000}\\
    &+\frac{b}{\sqrt{2}} \left[ \ket{100110}+\ket{101001}\right] +c\ket{011010},
    \end{aligned}
\end{equation}
where $a,b,c\in\mathbb{R}$ and $a^2+b^2+c^2 =1$. Evidently, we need all six qubits in the quantum computation of the first excited state. As with the ground state, we consider two different ans\"{a}tze, one consisting of two terms and a single-term one. Both are given in terms of three variational parameters and contain three and eight CNOT gates, respectively.

The two-term ansatz is given by
\begin{equation} \label{E12_eqn}
    \ket{\psi_\text{trial}}=\sin\chi\ U_1' \ket{000000} + \cos\chi\ U_2' \ket{000000}
\end{equation}
where
\begin{equation}
\begin{aligned} \label{eq:47}
    U_1' &= X_0 CX_{25} X_2 CX_{34} {CR_Y}_{23} (\phi) {R_Y}_2 (\theta) \\ U_2' &= X_1 X_2X_4
    \end{aligned}
\end{equation}
with variational parameters $\theta$, $\phi$, and $\chi$.
The unitaries $U_1'$ (containing 3 CNOT gates) and $U_2'$ (no CNOT gates) are implemented by these two circuits, respectively,
\[\Qcircuit @C=0.4em @R=0.5em {
    \lstick{\ket{0}_0} & \gate{X} & \qw & \qw & \qw & \qw & \qw \\
    \lstick{\ket{0}_2} & \gate{R_Y(\theta)} & \ctrl{1} & \gate{X} & \qw & \ctrl{3} & \qw \\
    \lstick{\ket{0}_3} & \gate{R_Y(\frac{\phi}{2})} & \targ & \gate{R_Y^\dagger(\frac{\phi}{2})} & \ctrl{1} & \qw & \qw \\
    \lstick{\ket{0}_4} & \qw & \qw & \qw & \targ & \qw & \qw \\
    \lstick{\ket{0}_5} & \qw & \qw & \qw & \qw & \targ & \qw
}
\hspace{1.2cm}
\Qcircuit @C=0.5em @R=0.5em {
    \lstick{\ket{0}_1} & \gate{X} & \qw \\
    \lstick{\ket{0}_2} & \gate{X} & \qw \\
    \lstick{\ket{0}_4} & \gate{X} & \qw
}\] 
We note that the two circuits are run independently in order to calculate the expectation value.
The single-term ansatz contains 8 CNOT gates and is implemented with this quantum circuit:

\[\Qcircuit @C=0.2em @R=0.5em {
    \lstick{\ket{0}_0} & \gate{X} & \qw & \qw & \targ & \qw & \qw & \qw & \qw & \qw & \qw & \ctrl{4} & \qw & \qw \\
    \lstick{\ket{0}_1} & \gate{R_Y(\frac{\chi}{2})} & \targ & \gate{R_Y^\dagger(\frac{\chi}{2})} & \ctrl{-1} & \ctrl{4} & \qw & \qw & \ctrl{4} & \qw & \qw & \qw & \qw & \qw \\
    \lstick{\ket{0}_2} & \qw & \qw & \qw & \qw & \qw & \targ & \ctrl{1} & \qw & \gate{X} & \qw & \qw & \qw & \qw \\
    \lstick{\ket{0}_3} & \gate{R_Y(\frac{\phi}{2})} & \qw & \qw & \qw & \qw & \qw & \targ & \qw & \gate{R_Y^\dagger(\frac{\phi}{2})} & \ctrl{1} & \qw & \qw & \qw \\
    \lstick{\ket{0}_4} & \qw & \qw & \qw & \qw & \qw & \qw & \qw & \qw & \qw & \targ & \targ & \gate{X} & \qw \\
    \lstick{\ket{0}_5} & \gate{R_Y(\theta)} & \ctrl{-4} & \qw & \qw & \targ & \ctrl{-3} & \gate{X} & \targ & \qw & \qw & \qw & \qw & \qw
}\]
in terms of the same three variational parameters.

\section{Error Mitigation}
\label{sec:5}

 While our ans\"{a}tze are virtually free of algorithmic errors, in the sense that minimizing the energy functional by classical algorithms leads to negligible errors, raw hardware runs lead to large errors, thus necessitating error mitigation. In this section, we describe the error mitigation techniques we used. In particular, we employ read-out (RO) error correction and an extrapolation scheme based on adding pairs of CNOT gates to our quantum circuits.

Read-out error corresponds to the error in the measurement process wherein there is a finite probability for the occurence of a classical bit flip. Let $p(0|1)$ and $p(1|0)$ denote these probabilities, where $p(0|1)$ is the probability that a true measurement outcome of $1$ is read-out as a $0$ instead, and similarly for $p(1|0)$. Following  \cite{Yeter-Aydeniz2019}, we define for qubit $i$ the symmetric and anti-symmetric likelihoods
\begin{equation}
p_i^{\pm}=p_i(0|1) \pm p_i(1|0)
\end{equation}
We account for RO errors using $p_i^{\pm}$ so that the corrected expectation value of a string of Pauli-$Z$ operators   $\langle Z_0^{x_0} Z_1^{x_1} \ldots Z_5^{x_5}\rangle$ (where each of the $x_i$ is either $0$ or $1$, corresponding to the absence or presence of $Z_i$ in the string, respectively) is
\begin{equation}
    \sum_{i_0,i_1,\ldots,i_5}  \Bigg\{P_{i_0,i_1,\ldots,i_5} \prod_{j\mid x_j = 1} \left[\frac{(-1)^{i_j} - p_j^{-}}{1-p_j^{+}}\right]\Bigg\}
\end{equation}
Here $P_{i_0,i_1,\ldots, i_5}$ is the probability obtained from measurement counts for the computational basis state $\ket{i_0,i_1,\ldots,i_5}$. We note that this error mitigation scheme assumes local, rather than correlated, RO errors.

\begin{figure}[h]
    \centering
    \includegraphics[scale=0.55]{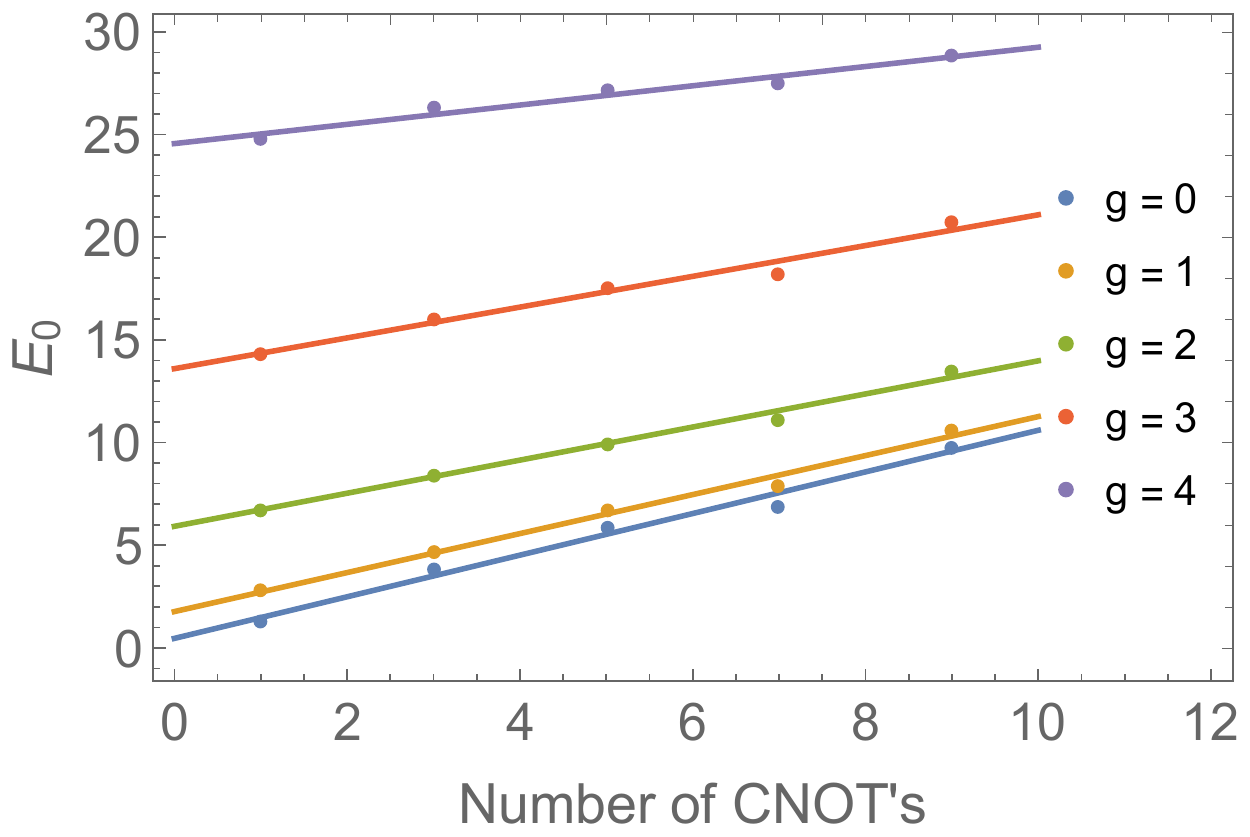}
    \caption{CNOT extrapolation for the minimized ground-state energy functional $E_0$ from the single-term ansatz. It is plotted as a function of the number of CNOT gates per CNOT in the original ansatz, for different values of the interaction strength $g$, $m_0=10$, and $\xi=0.7$.} 
    \label{E0_1_term_RE}
\end{figure}

\begin{figure}[h] 
    \centering
    \includegraphics[scale=0.55]{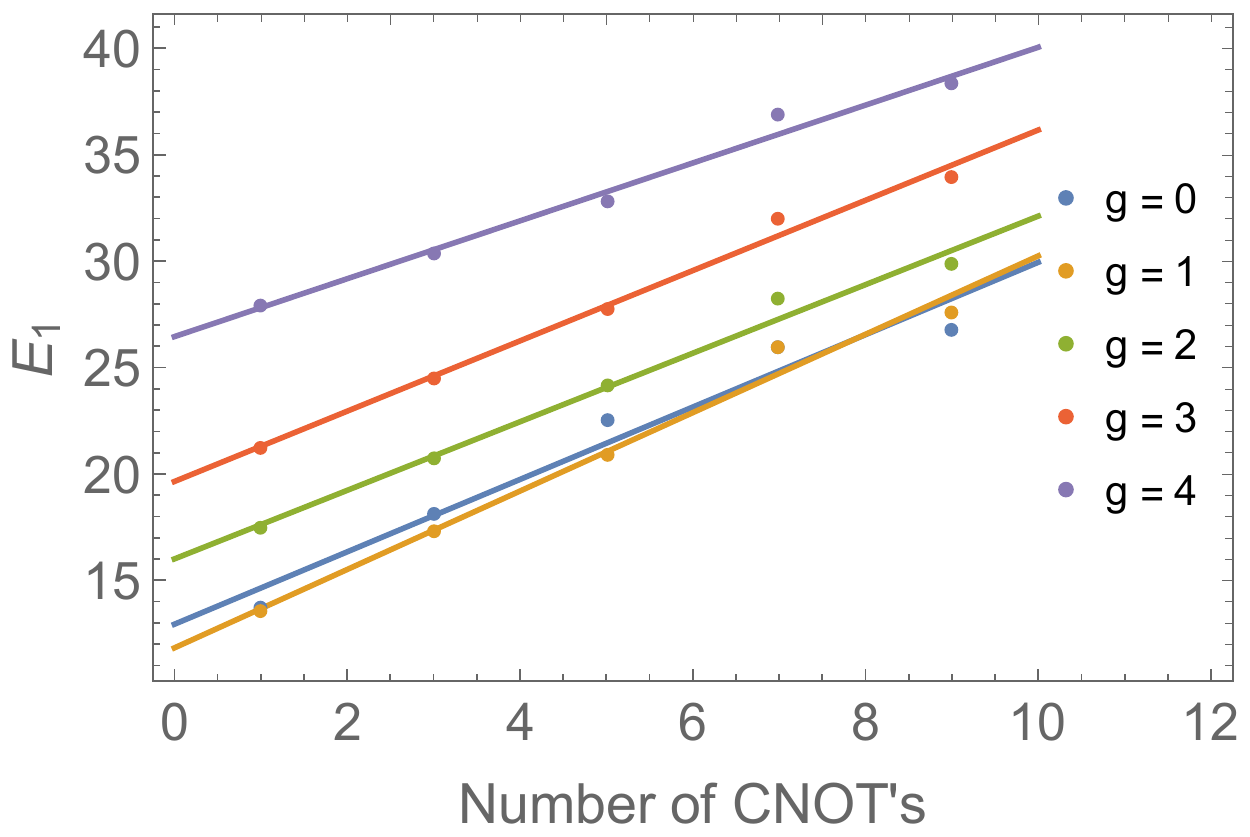}
    \caption{CNOT extrapolation for the minimized first-excited-state energy functional from the single-term ansatz. It is plotted as a function of the number of CNOT gates per CNOT in the original ansatz, for different values of the interaction strength $g$, $m_0=10$, and $\xi=0.7$.} 
    
    \label{E1_1_term_RE}
\end{figure}

The RO corrected data thus obtained still suffers from significant CNOT errors. In order to compensate for these errors, following \cite{Li2017} we employ an extrapolation scheme wherein we include additional pairs of CNOT gates for every CNOT gate in our ansatz. We take five such runs so that we go up to 9 CNOT gates per CNOT gate in the original ansatz. Applying a linear fit to these energies, we obtain the approximate zero-noise value of the energy by extrapolating to the 0-CNOT-gate case, as shown in Figs.\ \ref{E0_1_term_RE} and  \ref{E1_1_term_RE}.

In the next section, we present results obtained after applying read-out error correction and CNOT-extrapolation.

\section{Results}
\label{Results}
\begin{figure*}[t]
    \centering
    \subfigure{\includegraphics[width=0.4\textwidth]{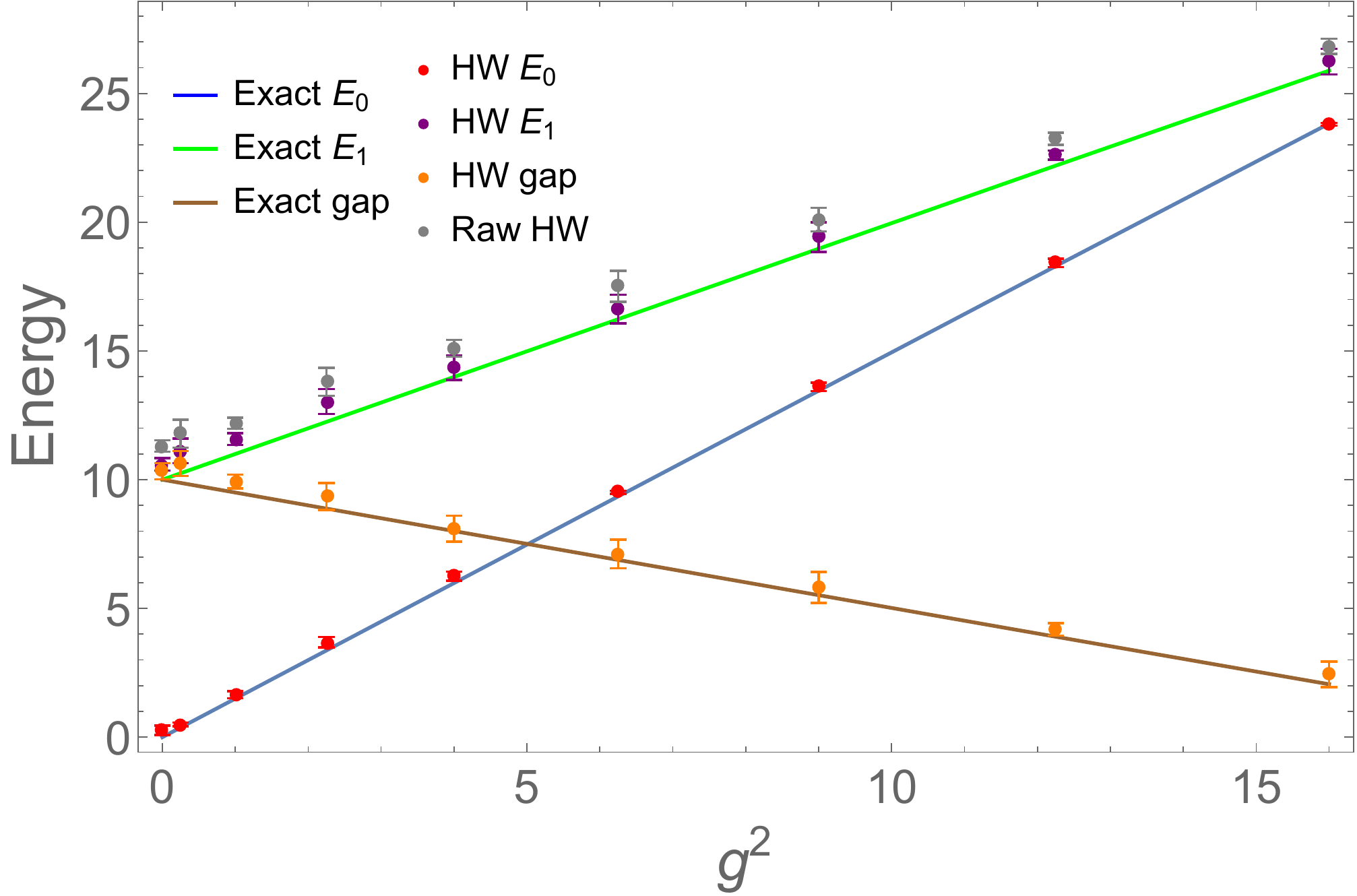}}
    \subfigure{\includegraphics[width=0.4\textwidth]{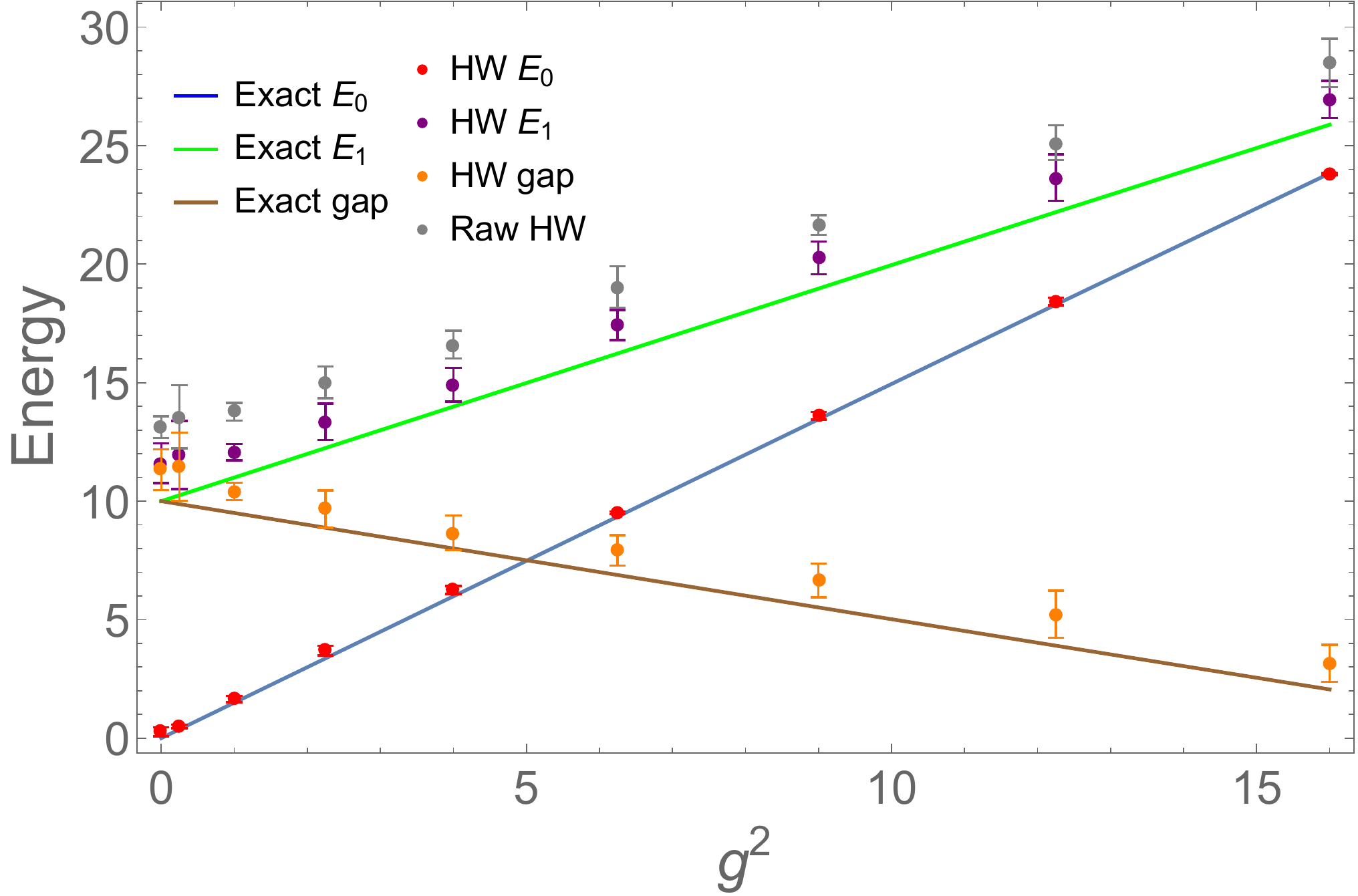}}
    \caption{$E$ vs $g^2$, where $E$ is the minimized energy functional for a given ansatz. Here $E_0$ is the ground-state energy from the 2-term ansatz and $E_1$ is the first-excited-state energy from the two-term (left) and single-term ansatz (right). We do not give ``raw" data for the $E_0$ results as it overlaps with CNOT-extrapolated data. The mass gap is $m=E_1-E_0$. ``HW" denotes Boeblingen hardware. The parameters are $m_0=10$ and $\xi=0.7$.}
    \label{E0_2terms}
\end{figure*}
\begin{figure*}[t]
    \centering
    \subfigure{\includegraphics[width=0.4\textwidth]{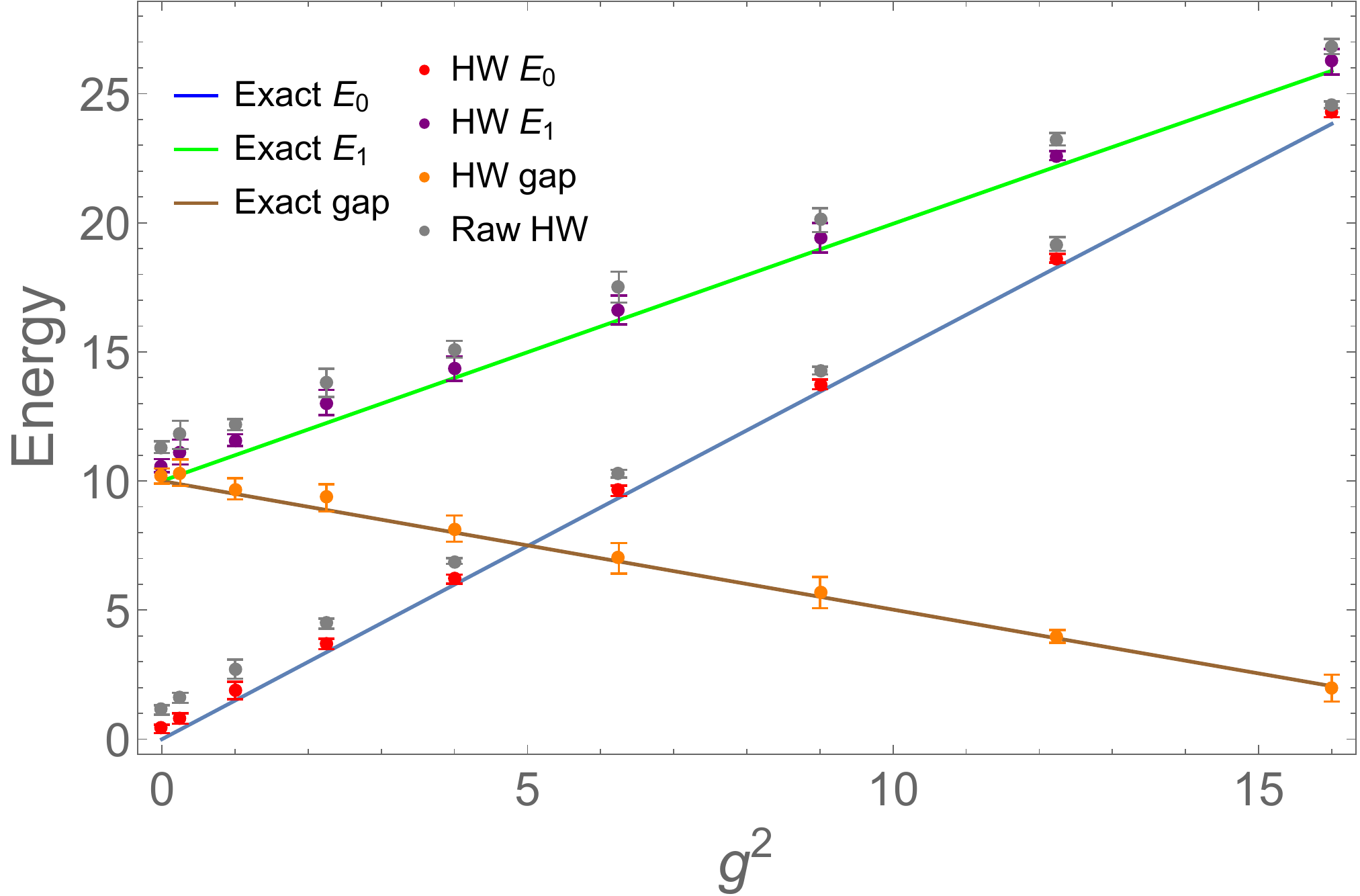}}
    \subfigure{\includegraphics[width=0.4\textwidth]{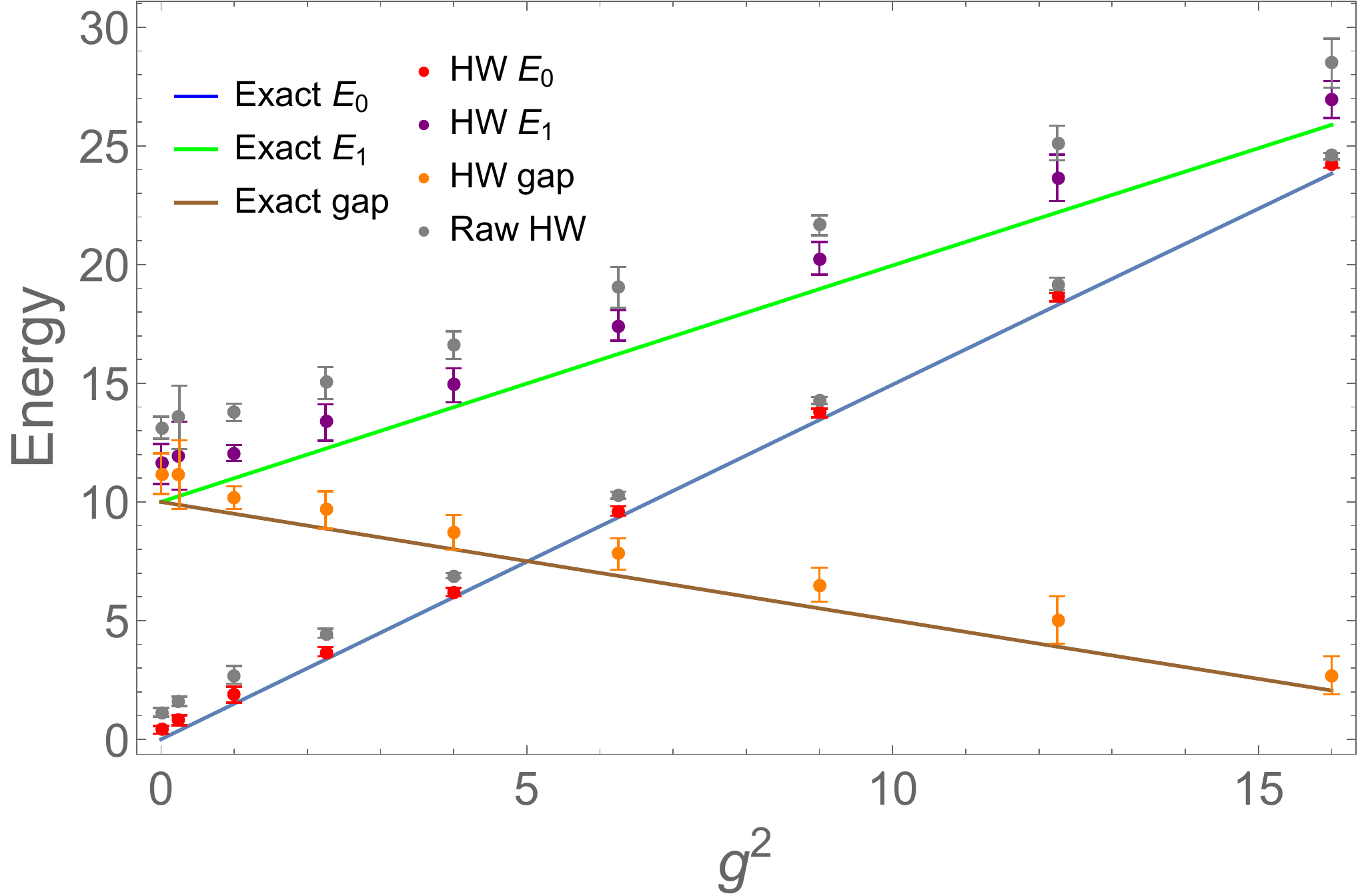}}
    \caption{$E$ vs $g^2$, where $E$ is the minimized energy functional for a given ansatz. Here $E_0$ is the ground-state energy from the single-term ansatz, and $E_1$ is the first-excited-state energy from the two-term (left) and 1-term ansatz (right). The mass gap is $m = E_1-E_0$. ``HW" denotes Boeblingen hardware. The parameters are $m_0=10$ and $\xi=0.7$.}
    \label{E0_1terms}
\end{figure*}
We now present the results obtained on the IBM Boeblingen quantum device for each of the four ans\"atze discussed in Section \ref{algorithms}. These are shown in Figs. \ref{E0_2terms} and \ref{E0_1terms}, where the ``raw'' data correspond to data obtained before CNOT extrapolation was applied but after RO error was accounted for. The error bars are statistical error bars obtained after several runs of the minimization on the device. 

In order to reduce computational overhead, we expressed two of the variational parameters in the excited state ansatz and one of the parameters in the ground state ansatz in terms of the others based on an approximation derived from noiseless simulations. We chose $\phi=\chi = \frac{\theta}{2}$ for all  of the circuits except for the first one (Eqs.\ \eqref{gd_2term} and \eqref{eq:41}), in which we chose $\phi=\pi-2\theta$. Thus, in all four ans\"atze, we minimized with respect to $\theta$ only. 

Unsurprisingly, we see the best agreement with exact results for the case where the ans\"{a}tze had two terms, i.e., the ones which contained fewer CNOT gates. For the ground state, the single-term ansatz did not introduce significant errors since only a couple of CNOT gates were added. However, switching to a single-term ansatz for the first excited state introduced significant errors as eight CNOT gates were required. In this case, linear extrapolation is not sufficient to match the accuracy obtained using the two-term first-excited-state ansatz.

\begin{figure}[h]
    \centering
    \subfigure{\includegraphics[width=0.4\textwidth]{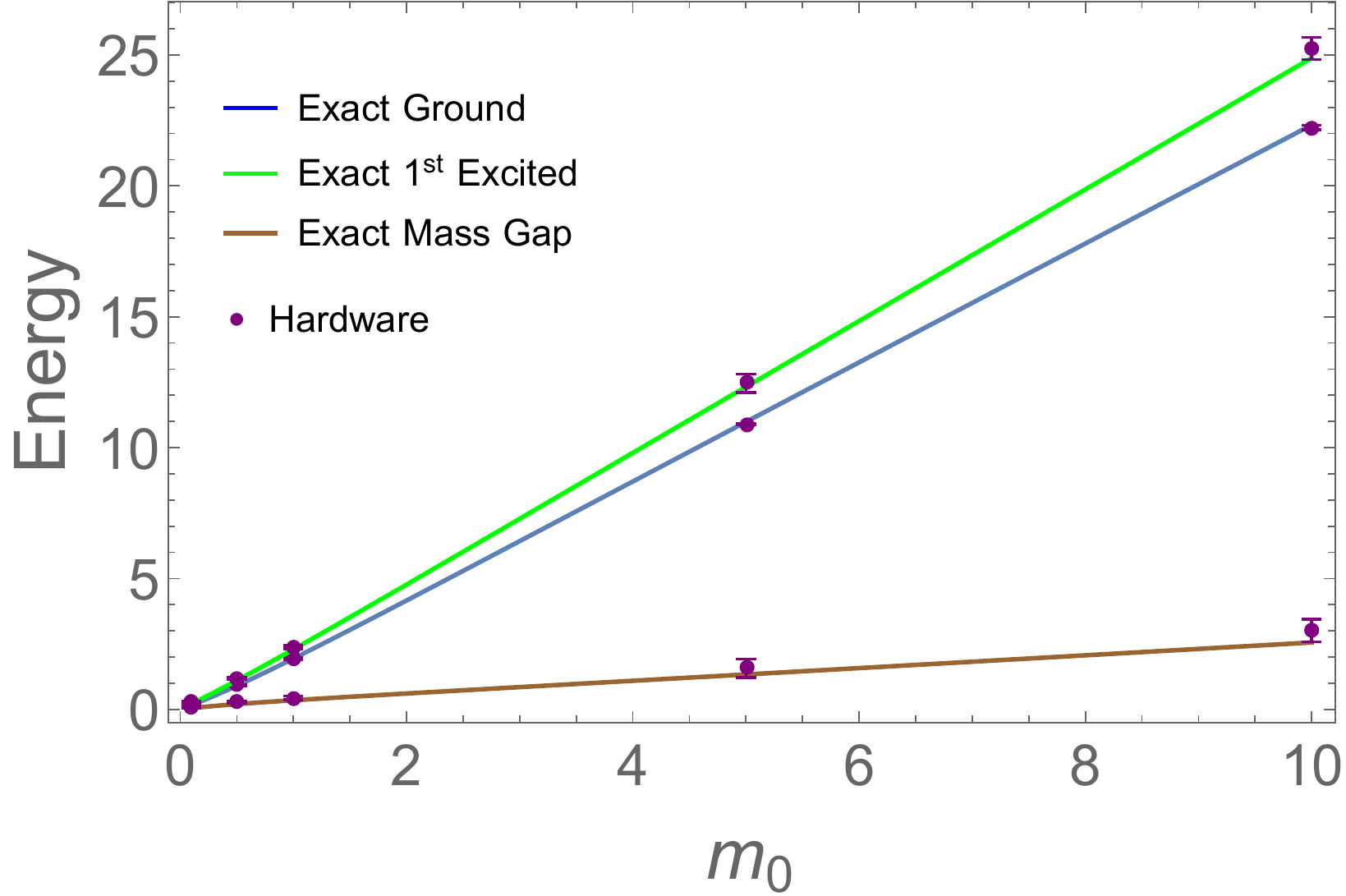}}
    \caption{Exact and hardware results demonstrating the vanishing of the mass gap as the sampling line, given by $m_0=\frac{2}{3}g^2$, approaches the origin. Here the energies and gap $m$ are plotted as functions of the bare mass $m_0$, with $\xi = 0.3$.}
    \label{Chiral_Results}
\end{figure}

As mentioned above, a physically interesting limit is the one where the mass scale vanishes exhibiting chiral symmetry. After discretization, this symmetry is only obtained approximately due to the addition of the Wilson term in the Hamiltonian (which vanishes in the continuum limit, but as a finite contribution on a finite lattice). We explored this limit with a quantum computation.
We did this by approaching the small $m_0$ region along a straight-line trajectory in the $m_0-g^2$ plane that passes through the origin and is steeper than the critical line. We also lowered the value of $\xi$ (the Wilson parameter)
to reduce the effect of the Wilson term on the chiral symmetry. Along this trajectory, we ran our quantum computation using the  two-term ans\"{a}tze for both $E_0$ and $E_1$. The results of these runs are given in Fig.\ \ref{Chiral_Results}.

\begin{figure}[h]
    \centering
    \subfigure{\includegraphics[width=0.4\textwidth]{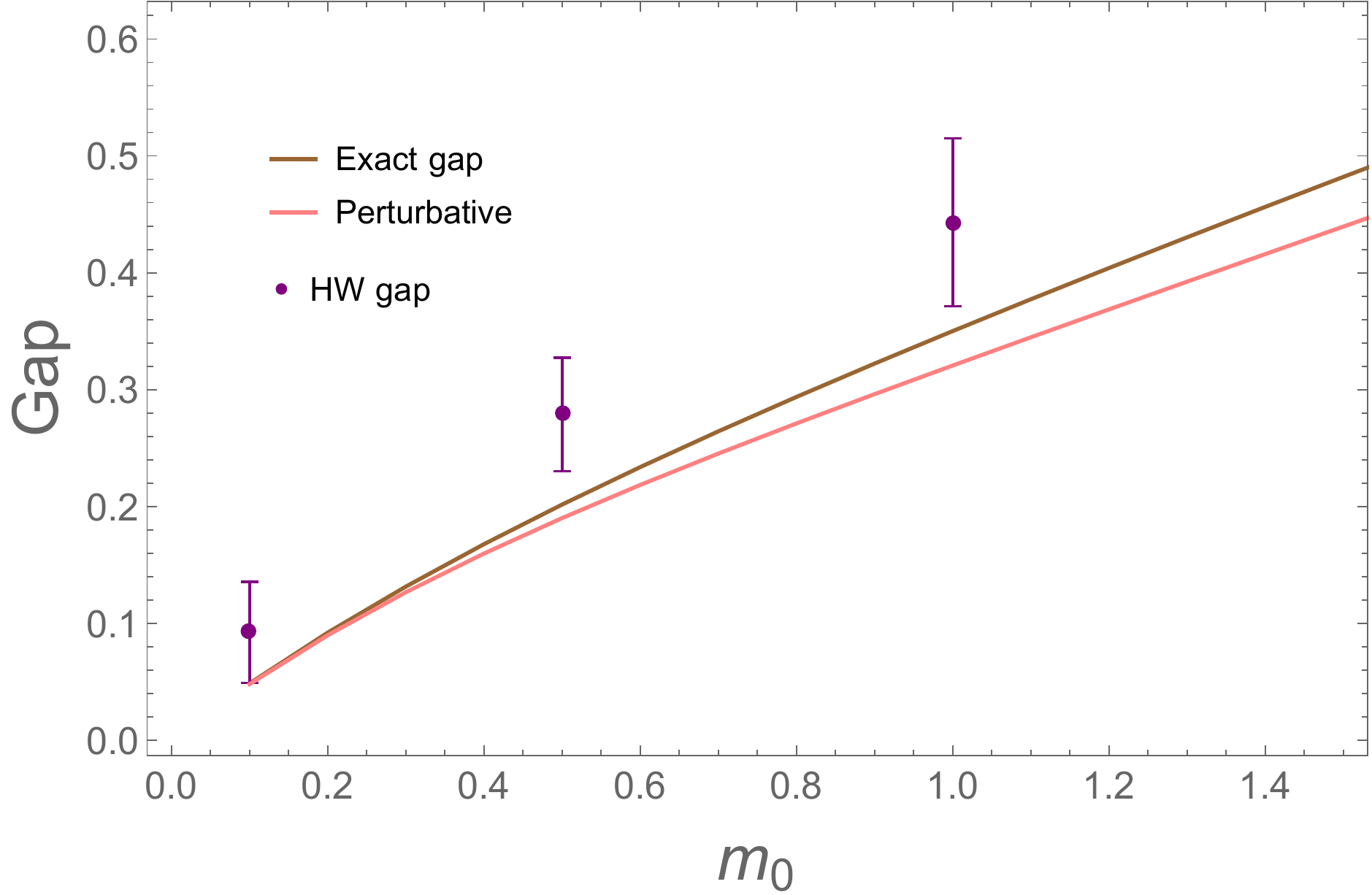}}
    \caption{Mass gap $m$ vs bare mass $m_0$ from Fig.\ \ref{Chiral_Results}, zoomed in near the origin to show where perturbation theory diverges from exact results.}
    \label{Chiral_Results_Pert}
\end{figure}

As seen in Fig.\ \ref{Chiral_Results_Pert}, where we zoom in to the low-bare-mass regime, the relative errors are rather large (while absolute error increases monotonically with $m_0$). 
It should also be noted that in this low-mass region, we find  that even the noiseless quantum simulations (using \verb|qasm_simulator|) produce results less accurate than those obtained via the straightforward minimizations of the energy functionals performed using standard classical techniques. It follows that further error mitigation will be necessary in this regime in order to perform a quantum computation exploring chiral symmetry in this model.

\section{Conclusion}
\label{sec:7}
In summary, we performed a hybrid classical-quantum calculation of energy levels of the massive Thirring model which is a relativistic fermionic quantum field theory (QFT) in one spatial dimension. To this end, we discretized the spatial coordinate while keeping time continuous. This allowed the use of Hamiltonian techniques rather than the classical Lagrangian approach. We used the Jordan-Wigner transformation to map our model onto qubits. For the trial state needed for the variational principle, we employed ans\"atze that were more general than those used in the Unitary Couple Cluster (UCC) method which is widely used in quantum chemistry. Some of our ans\"atze were superpositions of wavefunctions requiring two unitaries to specify them. 
This handed over portions of the calculation to the practically error-free classical part of the code, mitigating some of the errors from NISQ quantum hardware. This also reduced the computational complexity of the algorithm in terms of the number of qubits required. We ran the quantum computation on the IBM Q device Boeblingen. To deal with device errors, we implemented Readout error correction and also a CNOT-extrapolation scheme. With error mitigation, we showed the results of the quantum computation to be in agreement with those obtained by exact classical computations. Moreover, we approached the chiral-symmetry limit (vanishing mass scale) through our quantum computation, and showed that the ground and the first-excited states approached degeneracy as we let the bare mass parameter $m_0 \rightarrow 0$ while staying within the physically relevant regime above the critical line.

It would be interesting to extend our approach to more complicated relativistic fermionic quantum field theories (higher dimensions, including of gauge symmetry, etc.) and benchmark NISQ quantum hardware by determining the maximum role the quantum algorithm can play within a hybrid algorithm for a given (acceptable) error level.

\acknowledgments
This work was supported as part of the ASCR Quantum Testbed Pathfinder Program at Oak Ridge National Laboratory under FWP \#ERKJ332. Part of this research used quantum computing system resources of the Oak Ridge Leadership Computing Facility, which is a DOE Office of Science User Facility supported under Contract DE-AC05-00OR22725. Oak Ridge National Laboratory manages access to the IBM Q System as part of the IBM Q Network.

\appendix
\section{Details of the Quantum Computation}
\label{app:a}

Here we provide some details of the procedure we used to reduce the number of Pauli strings in the Hamiltonian in order to reduce the errors due to NISQ quantum hardware to an acceptable level.

It is imperative that we reduce the total number of Pauli strings $H_l$ in the Hamiltonian. After a choice of trial function is made, we need to identify terms of the form $\bra{0}U_m^{\dagger}H_{l}U_n\ket{0}$ (see Eq.\ \eqref{expectation_terms}) that are related to each other.

For example, for the unitary $U_1$ appearing in the ground-state ansatz \eqref{gd_2term}, and involves qubits 3 and 4, it is easy to see that
\begin{equation}
    Y_3Y_4 U_1\ket{0} = -X_3X_4 U_1\ket{0}
\end{equation}
so terms in the Hamiltonian containing the strings $Y_3Y_4$ and $X_3X_4$ can be combined.

Another example is the string $Z_0Z_2$ in the Hamiltonian. It is easy to see that each term in the ansatz \eqref{gd_2term} is an eigenstate of this string,
\begin{equation}
    Z_0Z_2 U_1 \ket{0} = U_1 \ket{0} \ , \ \ Z_0Z_2 U_2 \ket{0} = - U_2 \ket{0}
\end{equation}
In each case, $Z_0Z_2$ can be replaced by its eigenvalue.

Such steps reduce the number of Pauli strings to 12, which is a manageable length for NISQ quantum hardware. The number of different quantum circuits needed is actually 4. This is because several of the 12 terms can be implemented with the same quantum circuit. For example, the terms $X_0Y_1Z_2$ and $X_0Y_1Z_5$ need a circuit for the common string $X_0Y_1$, because $Z_2$ and $Z_5$ are already diagonal in the computational basis. The string $X_0Y_1$ can be diagonalized in the standard way using Eq.\ \eqref{eq:43}.

For terms $\bra{0}U_m^{\dagger}H_{l}U_n\ket{0}$ with $m\ne n$, which only occurs when we go beyond the standard UCC method and consider trial states consisting of two (or more) terms, such as the ansatz \eqref{gd_2term}, we ought to diagonalize $U_m^{\dagger}H_{l}U_n$ and not just $H_l$ by itself.

For example, consider the term $\bra{0} U_1^\dagger H_l U_2 \ket{0}$ with $H_l = X_2X_5$, which appears when we employ the ansatz \eqref{gd_2term} for the ground state. Since $H_l = U_2$, this transition amplitude simplifies to
\begin{equation}
    \bra{0} U_1^\dagger X_2 X_5 U_2 \ket{0} = \bra{0} e^{i\frac{\phi}{2} X_3 Y_4} \ket{0}
\end{equation}
where we used \eqref{eq:41}. This can be easily diagonalized using \eqref{eq:43}. We obtain
\begin{equation}
    \bra{0} U_1^\dagger X_2 X_5 U_2 \ket{0} = \bra{0}  \mathcal{U}^\dagger  e^{i\frac{\phi}{2} Z_3 Z_4} \mathcal{U} \ket{0}
\end{equation}
where $\mathcal{U} = \mathcal{Y}_4 \bm{H}_3$.

A slightly more complicated example is a term of the same form but with $H_l = X_2 X_3 X_4 X_5$. It simplifies to
\begin{equation}
    \bra{0} U_1^\dagger X_2 X_3 X_4 X_5 U_2 \ket{0} = \bra{0} e^{i\frac{\phi}{2} X_3 Y_4} X_3 X_4 \ket{0}
\end{equation}
In this case, before we use Eq.\ \eqref{eq:43}, we observe that $X^+ \ket{0} = 0$, where $X^\pm = X\pm iY$, therefore we can replace $X_4$ by $-i Y_4$. After using \eqref{gd_2term}, we obtain
\begin{equation}
    \bra{0} U_1^\dagger X_2 X_3 X_4 X_5 U_2 \ket{0} = \bra{0}  \mathcal{U}^\dagger  e^{i\frac{\phi}{2} Z_3 Z_4} Z_3 Z_4 \mathcal{U} \ket{0}
\end{equation}
In summary, the above techniques dramatically reduce the number of terms in the Hamiltonian (e.g., for $N=3$, we have 166 terms) to a handful of quantum circuits, thus making the quantum computation manageable by NISQ quantum hardware.

\bibliographystyle{unsrt}
\bibliography{main}
\end{document}